# Giant and Tunable Bosonic Quantum Interference Induced by Two-Dimensional Metals

Kunyan Zhang[1,2], Rinu Abraham Maniyara[3], Yuanxi Wang[4], Arpit Jain[3], Maxwell T. Wetherington[5], Thuc T. Mai[6], Chengye Dong[3], Timothy Bowen[3], Ke Wang[5], Slava V. Rotkin[5,7], Angela R. Hight Walker[6], Vincent H. Crespi[8], Joshua Robinson[3,9], Shengxi Huang[1,2,*]

[1] Department of Electrical Engineering, The Pennsylvania State University, University Park, PA 16802, USA
[2] Department of Electrical and Computer Engineering, Rice University, Houston, TX 77005, USA
[3] Department of Materials Science and Engineering, The Pennsylvania State University, University Park, PA 16802, USA
[4] Department of Physics, University of North Texas, Denton, TX 76203, USA
[5] Materials Research Institute, The Pennsylvania State University, University Park, PA 16802, USA
[6] Quantum Measurements Division, Physical Measurement Laboratory, National Institute of Standards and Technology, Gaithersburg, MD 20899, USA
[7] Department of Engineering Science and Mechanics, The Pennsylvania State University, University Park, PA 16802, USA
[8] Department of Physics, The Pennsylvania State University, University Park, PA 16802, USA
[9] Two-Dimensional Crystal Consortium and Materials Research Institute, The Pennsylvania State University, University Park, PA 16802, USA
*Corresponding author: Shengxi Huang (shengxi.huang@rice.edu)

**Abstract**

Harnessing quantum interference among bosons provides significant opportunities as bosons often carry longer coherence time than fermions[1]. As an example of quantum interference, Fano resonance involving phonons or photons describes the coupling between discrete and continuous states, signified by an asymmetric spectral lineshape[2]. Utilizing photon-based Fano resonance, molecule sensing with ultra-high sensitivity and ultrafast optical switching has been realized[3,4]. However, phonon-based Fano resonance, which would expand the application space to a vaster regime, has been less exploited because of the weak coupling between discrete phonons with continuous states such as electronic continuum. In this work, we report the discovery of giant phonon-based Fano resonance in a graphene/2D Ag/SiC heterostructure. The Fano asymmetry, being proportional to the coupling strength, exceeds prior reports by two orders of magnitude[5-7]. This Fano asymmetry arises from simultaneous frequency and lifetime matching between discrete and continuous phonons of SiC. The introduction of 2D Ag layers restructures SiC at the interface and facilitates resonant scattering to further enhance the Fano asymmetry, which is not achievable with conventional Ag thin films. With these unique properties, we demonstrated that the phonon-based Fano resonance can be used for ultrasensitive molecule detection at the single-molecule level. Our work highlights strong Fano resonance in the phononic system, opening avenues for engineering quantum interference based on bosons. Further, our findings provide opportunities for advancing phonon-related applications, including biochemical sensing, quantum transduction, and superconductor-based quantum computing.

Quantum interference lies at the heart of quantum information science and technology[1], finding applications in precise quantum metrology[8], single-molecule detection[9], and quantum computation and cryptography[10]. One intriguing manifestation of quantum interference is Fano resonance, which arises from the interference between a discrete state $\varphi$ and a continuum of states $\psi_E$, as depicted in Fig. 1a, characterized by a coupling matrix element $\langle\psi_E|H|\varphi\rangle = V_E$. Here, $|V_E|^2$ serves as a measure of the strength of intereference.[2] Fano resonance is signified by an asymmetric Breit-Wigner-Fano (BWF) lineshape

$$I(\omega) = I_0 \frac{[1 + (\omega - \omega_0)/(q\Gamma_0/2)]^2}{1 + [(\omega - \omega_0)/(\Gamma_0/2)]^2} \quad (1)$$

in which $I_0$, $\omega_0$, and $\Gamma_0$ denote the intensity, frequency, and linewidth of the modified discrete state, respectively. The asymmetry parameter $|1/q|$ quantifies the level of Fano resonance, being proportional to $|V_E|$ (Fig. 1b and Supplementary Note 1)[2]. In photonic systems, the extreme antiresonance where $|1/q|$ approaches $\infty$ (Fig. 1a, purple curve) can be achieved through strong interference between two photonic oscillators with the interference strength surpassing the damping of the discrete oscillator. This strong Fano resonance has been observed in the scattering, transmission, or reflection spectra of photonic systems, underlining a plethora of applications including optical switching, lasing, and biosensing[3]. Attaining a high $|1/q|$ is critical for achieving superior performance in these applications. For instance, in biosensing, large $|1/q|$ ensures a significant spectral change with a slight change of $1/q$, offering high sensitivity and selectivity for molecule detection[3,4]. Moreover, when the energy of the discrete state equals that of the continuum, electromagnetically induced transparency (EIT) is observed, which enables the generation of slow light for effective quantum transduction[11].

While photonic systems have showcased the capacity to attain strong Fano resonance, phonon-based Fano resonance tends to exhibit weaker interference, thereby hindering the development of phonon-related devices, such as infrared imaging and Raman-based biosensors, superconductor-based quantum computers, and thermoelectric devices for energy harvesting. Investigation of phonon-based Fano resonance reveals that the continuous state often involves an electronic continuum with an intrinsically different lifetime compared to phonons. Specifically, electron-hole pairs exhibit a lifetime in the nanosecond range[12,13], while optical phonons possess a picosecond lifetime (Fig. 1c). The significant difference in lifetimes, spanning three orders of magnitude, prevents the effective interference between the electronic continuum and phonons. The lifetime of the modified discrete state $\tau$ is further inversely related to the strength of interference through $1/\tau \propto |V_E|^2$. Consequently, such Fano resonance typically yields a $|1/q|$ value (proportional to $|V_E|$) smaller than 1 as evidenced in materials like graphene[5,6], carbon nanotube (CNT)[14,15], Weyl semimetals[16,17], and semiconductors[7,18-20] (Fig. 1d). An enhanced Fano asymmetry can be achieved when interfering states possess lifetimes of similar magnitudes[21,22], i.e., both are supported by phonons. In this context, stronger Fano resonance has been observed in ferroelectric perovskite $KTa_{1-x}Nb_xO_3$[23,24] (Fig. 1d), which is attributed to the coupling between the discrete Ta-O vibrational state and the continuous Nb-O vibrational states. However, the quest for a larger $|1/q|$ represented by the total antiresonance persists in various applications, which is vital for pushing the boundaries of phonon-based devices and unlocking their full potential in quantum technologies.

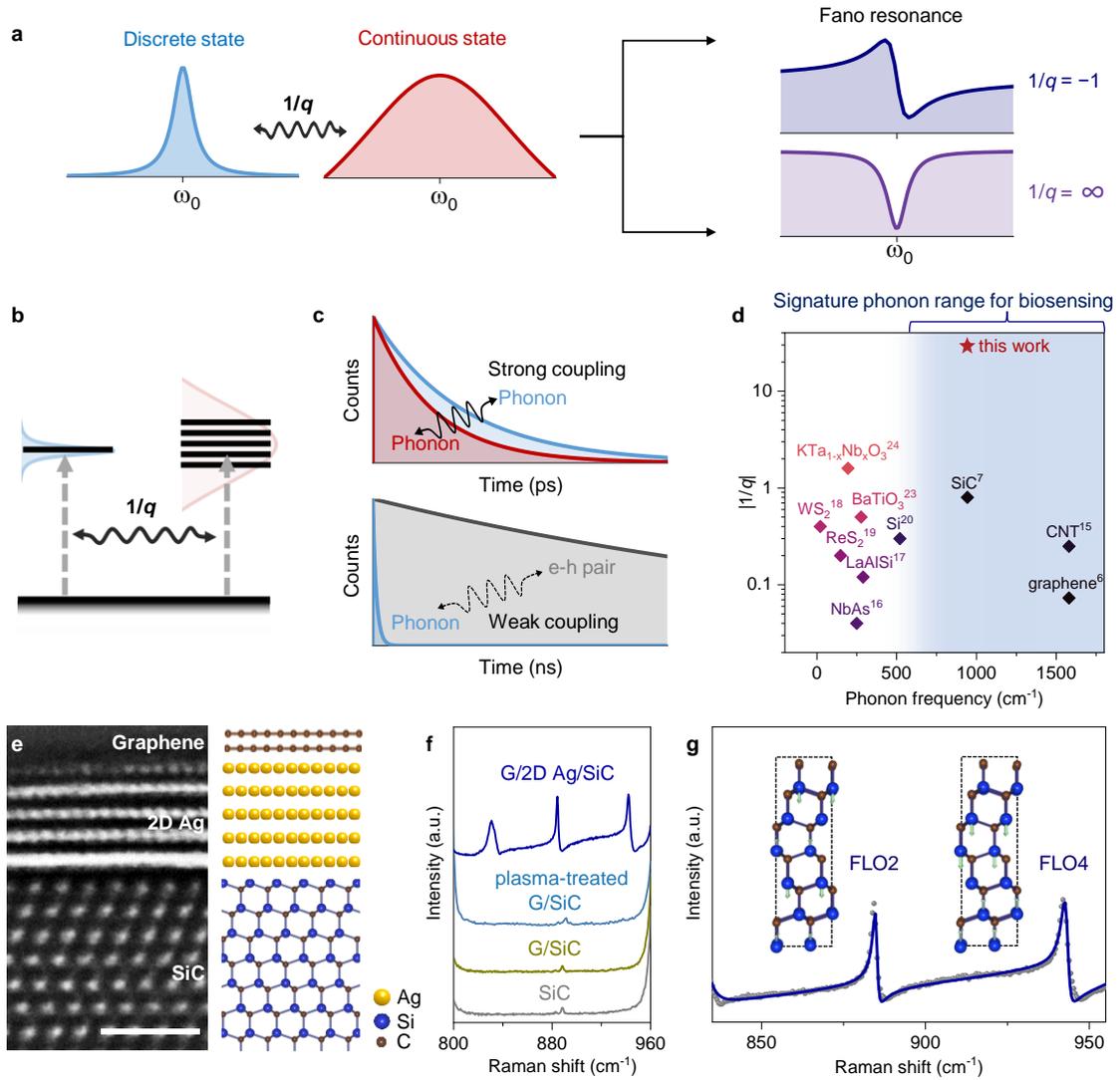

**Fig. 1**. Fano resonance in graphene/2D Ag/SiC. **a**, The spectra of dressed phonon as a result of the discrete state interfering with continuous states. **b**, Illustration of frequency (energy) matching between discrete and continuous states. **c**, Illustration of lifetime matching where the phonon lifetime is in picoseconds while the electron-hole pair lifetime is in nanoseconds, hindering the interference between phonons and electronic excitations. **d**, The reported $|1/q|$ as a function of phonon frequency for different materials where Raman-active phonons serve as the discrete state, such as graphene[6], CNT[15], Si[20], SiC[7], NbAs[16], LaAlSi[17], WS$_2$[18], ReS$_2$[19], BaTiO$_3$[23], KTa$_x$Nb$_{1-x}$O$_3$[24]. The $|1/q|$ in the figure is the highest value for each material best to our knowledge. **e**, Cross-sectional STEM images and atomic structure of graphene/2D Ag/SiC. Scale bar: 1 nm. **f**, Raman spectra of bare SiC, epitaxial graphene on SiC (G/SiC), oxygen plasma treated epitaxial graphene on SiC (plasma-treated G/SiC), and graphene/2D Ag/SiC (G/2D Ag/SiC) under the 532 nm excitation. The spectra are shifted along the y-axis. **g**, The enlarged FLO2 and FLO4 modes. The gray dots are the measured data, and the blue curve is the fitted spectrum.

In this work, we achieved a giant phonon-based Fano resonance characterized by a total antiresonance pattern and a $|1/q|$ value two orders of magnitude higher than most prior reports. This was accomplished in a graphene/2D Ag/SiC heterostructure (Fig. 1d). The strong Fano resonance is made possible by the intricate interplay between SiC and 2D Ag: first, both the continuous and discrete states are comprised of phonons of SiC, allowing for the frequency and temporal overlaps between the states; second, the presence of 2D Ag provides an essential intermediate state for resonant Raman scattering, which further enhances Fano resonance. These crucial effects are unique to this system and cannot be replicated by neutron-irradiated SiC, Ag thin film on SiC, or bare SiC, due to the absence of the specific transition pathways introduced by 2D Ag or the distinct atomic configurations in SiC, or both. Additionally, we demonstrated that the intensity and asymmetry of the Fano modes are extremely sensitive to the introduction of Rhodamin 6G molecules at the single molecule level: the intensity is modified by 43%, and the asymmetry factor $|1/q|$ changes as much as 87% compared to pristine graphene/2D Ag/SiC heterostructure. The phonon-based Fano resonance in our graphene/2D Ag/SiC is situated within the spectral range of signature phonons of biosystems (500 cm$^{-1}$ to 1700 cm$^{-1}$, Fig. 1d)[25]. Therefore, when extending the detection to complex biosystems, this advancement will enable the application of Fano resonance in phonon-based bioimaging, biosensing, and bioenergy harvesting.

**Giant Fano resonance at 2D Ag/SiC interface**

The graphene/2D Ag/SiC heterostructure was synthesized by confinement heteroepitaxy[26], a technique where a few layers of Ag atoms were intercalated between graphene and 6H-SiC as shown by the cross-sectional scanning transmission electron microscopy (STEM) image in Fig. 1e. The intercalation of 2D Ag leads to the activation of several phonon modes, which are not observed in oxygen plasma treated epitaxial graphene on SiC (plasma-treated G/SiC), epitaxial graphene on SiC (G/SiC), and bare SiC (Fig. 1f, full spectra in Fig. S1). These asymmetric phonons at 885 cm$^{-1}$ and 943 cm$^{-1}$ in Fig. 1g are assigned to the folded longitudinal optical (FLO) phonons of SiC based on first-principles calculations (Fig. S2 and Supplementary Note 2). The BWF fitting of the FLO modes yields $1/q$ of −0.4 and −0.5 for the FLO2 and FLO4 modes, respectively, validating the occurrence of Fano resonance in the FLO phonons of SiC following the metal intercalation process.

Our first-principles calculations predict a total of five FLO modes in 6H-SiC, designated as FLO1 to FLO5 (Fig. S2) for ease of reference. The phonon symmetry of the FLO1 to FLO5 modes is $B_1$, $A_1$, $A_1$, $B_1$, and $B_1$, respectively, in which the $A_1$ phonon is Raman-active while the $B_1$ phonon is Raman-inactive. This is consistent with the observation of weak FLO2 and FLO3 modes in bare SiC, G/SiC, and plasma-treated G/SiC (Fig. 1f). The polarization-resolved Raman measurement also confirms the $A_1$ symmetry of the FLO2 phonon (Fig. S3 and Supplementary Note 3). The appearance of the FLO4 mode after metal intercalation suggests the existence of new scattering states or breaking of crystal symmetry.

In addition to the spectrum involving the FLO2 and FLO4 modes (FLO2-4) discussed above, subsequent measurements consistently reveal two distinct patterns: one involving the FLO2 and FLO5 modes (FLO2-5) and the other involving the FLO3 and FLO4 modes (FLO3-4) as shown in Fig. 2a. Of particular interest is the unprecedented Fano resonance observed in the FLO4 mode, wherein a $1/q = -30.8$ is achieved in the FLO3-4 pattern (Fig. 2b, purple curve). This value surpasses commonly reported values for phonon-related Fano resonance by two orders of magnitude (Fig. 1d). The depth profiles of these Fano patterns reveal that the Fano resonance is localized solely at the interface between 2D Ag and SiC (Fig. S4), where the atomic structure of SiC is modified by metal intercalation[26,27]. SiC exhibits a typical terrace morphology with a terrace size of 5-10 μm and a step height of 5-10 nm as shown by the atomic force microscopy (AFM) images (Fig. S5). The terrace step edge is also visible in the optical image (Fig. 2c, dashed line). The observed Fano pattern is distinctly distributed across the SiC terrace as shown by the alternating FLO2-4 and FLO2-5 patterns along adjacent terraces (Figs. 2c-d and Fig. S6). Similarly, we observed the presence of FLO2-4 and FLO3-4 patterns along the SiC steps (Figs. 2e-f). The spatial distribution of these Fano patterns suggests that they are related to the SiC surface terminations, which naturally vary across different terraces.

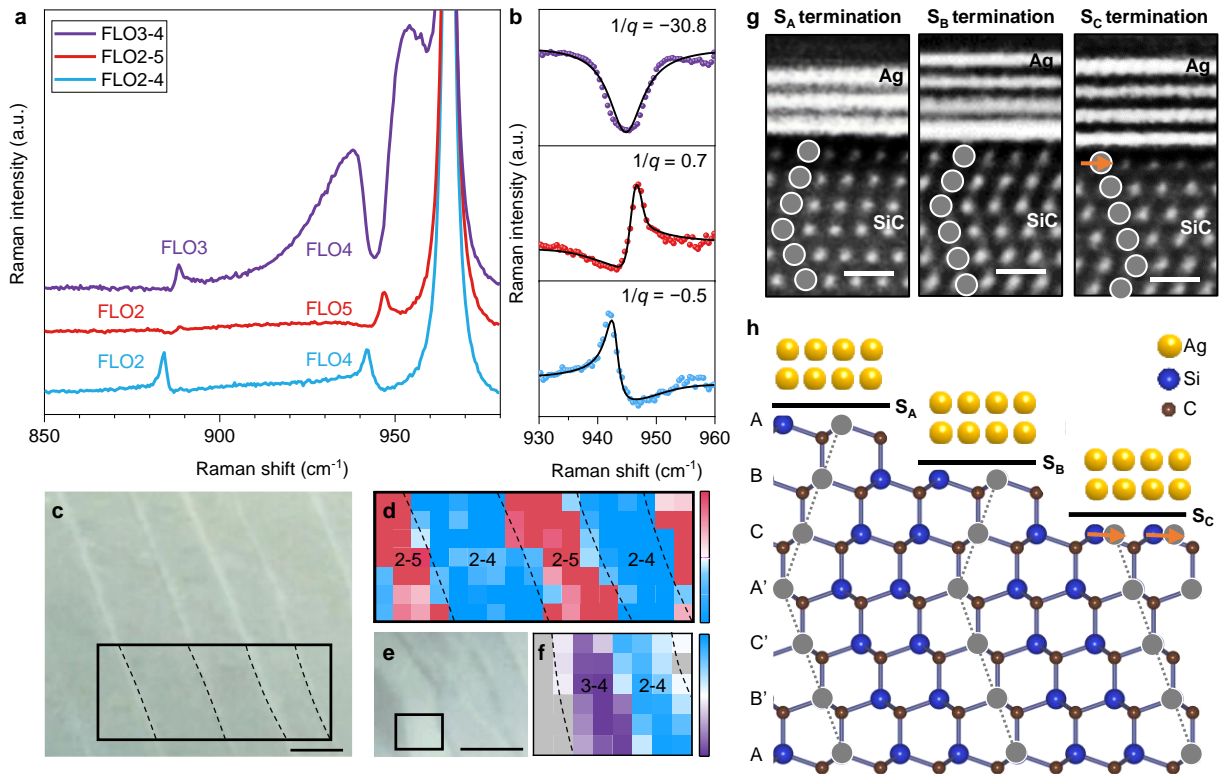

**Fig. 2**. Three patterns of Fano resonance determined by SiC terminations. **a**, Raman spectra of different Fano patterns. The spectra are shifted along the y-axis. **b**, The least-square fitting of the FLO4 or FLO5 mode in **a** using BWF function after background subtraction. **c**, The optical image of graphene/2D Ag/SiC, sample #1. The scale bar is 5 μm. **d**, Mapping of the FLO4 intensity in a FLO2-4 pattern (blue) and the FLO5 intensity in a FLO2-5 pattern (red) for the box area in **c**. **e**, The optical image of graphene/2D Ag/SiC, sample #2. The scale bar is 5 μm. **f**, Mapping of the FLO4 intensity in a FLO2-4 pattern (blue) and in a FLO3-4 pattern (purple) for the box area in **e**. **g**, STEM images of graphene/2D Ag/SiC with different SiC terminations. The gray dots label the periodicity of Si. **h**, Illustration of 6H-SiC. The black curve across the surface represents the terminations at different planes. The gray dots and the dotted lines denote the periodicity of Si. The orange arrows represent the reconfiguration of the Si at the $S_C$ termination.

We attribute these Fano patterns to the kinetic and thermodynamic preference of SiC terminations, respectively. As shown in Fig. 2h, 6H-SiC exhibits a vertical periodicity of six Si layers labeled as ABCA'C'B'. The local bonding environment of A, B, and C planes resembles that of A', C', and B' planes, resulting in three surface terminations denoted as $S_A$, $S_B$, or $S_C$. Upon graphene formation, the SiC surface undergoes kinetically driven reconstruction, where Si retracts at varying rates depending on the terminations[28]. Specifically, $S_C$ retracts faster, leading to a periodically stepped surface with $S_A$ and $S_B$, thereby giving rise to the alternating FLO2-4 and FLO2-5 patterns observed across steps. Meanwhile, the coexistence of both FLO2-4 and FLO3-4 patterns (Fig. 2f and Fig. S6) is likely thermodynamically driven, since $S_A$ and $S_C$ are more energetically favorable than $S_B$ by 13 meV per lateral unit cell (i.e. per surface Si) according to our density functional theory (DFT) calculations. The correlation between three Fano patterns and three SiC terminations is further corroborated by the calculated phonon intensity (Fig. S7). Therefore, we have confirmed that the Fano patterns are determined by the surface termination of SiC, where the vibrations of top Si are enhanced by 2D metal. Based on the STEM images, we can identify two types of defects in SiC after metal intercalation: Si vacancies and stacking faults. Si sublimation that forms graphene can induce Si vacancies on the SiC surface, as evidenced by the reduced brightness of the topmost Si layer compared to the underlying layers (Fig. 2g). In some instances, Ag atoms occupy these vacancies as shown in Fig. S8. Similar metal substitution at Si vacancies has been reported in graphene/2D Ga/SiC heterostructure[29]. Additionally, stacking faults are present in SiC (Figs. 2g-h, orange arrow, also in Fig. S8), where the top Si layer exhibits a lateral translation that resembles the stacking structure of 3C-SiC instead of 6H-SiC. These SiC defects can act as scattering centers, potentially activating the Raman-inactive FLO modes and broadening the LO phonon through the phonon confinement effect, which will be discussed in detail later.

**Origin of scattering pathways**

To understand the underlying mechanism behind the remarkable Fano resonance, we look into the origin of scattering pathways. The continuum associated with the FLO2-4 pattern spans the spectral range from 800 cm$^{-1}$ to 965 cm$^{-1}$, while it is confined within 900 cm$^{-1}$ to 965 cm$^{-1}$ for the FLO3-4 pattern. The distinct lineshape of the continuum is consistent with the confined LO phonon of SiC modified by the phonon confinement effect. Such phonon confinement effect has been observed in neutron- and ion-irradiated SiC (Fig. S9)[30,31]. In the case of our graphene/2D metal/SiC, a similar effect is induced by metal intercalation[28] that creates atomic vacancies or stacking faults on the SiC surface, which serve as scattering centers for phonons. Therefore, the Raman intensity in the phonon confinement model is expressed as an integration of phonon dispersion (Fig. 3a)[30,32,33]

$$I(\omega) = \int \frac{\exp(-k^2 D^2/4)}{[\omega - \omega(k)]^2 + (\Gamma_{LO}/2)^2} d^3k \qquad (2)$$

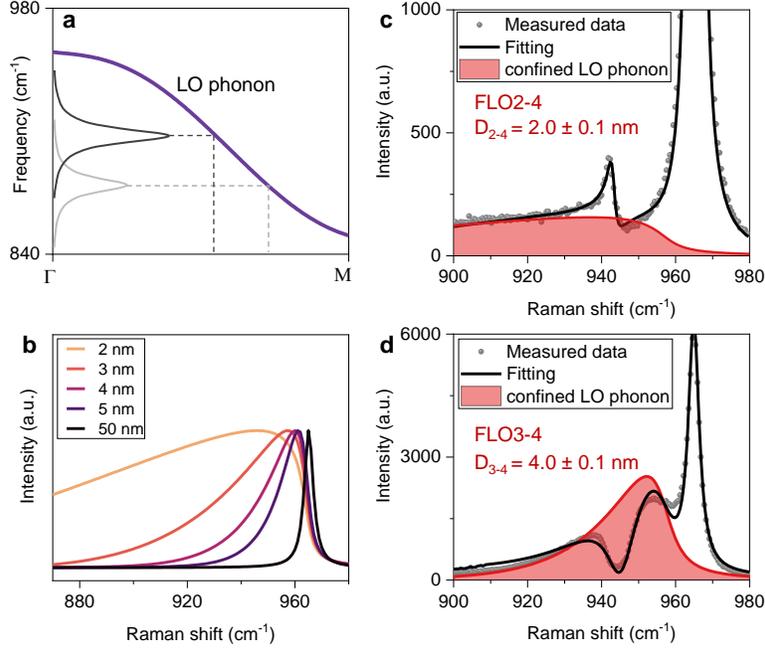

**Fig. 3.** Phonon confinement analysis of the continuous states. **a**, Dispersion of the LO phonon along the Γ-M direction. The gray spectra illustrate that phonons away from the Γ point contribute to the overall intensity. **b**, The constructed spectra using the phonon confinement model show the broadening of the confined LO phonon with different correlation lengths $D$. Fitting of the scattering background using the phonon confinement model for **c**, FLO2-4 and **d**, FLO3-4 patterns.

$\omega(k)$ represents the phonon dispersion, $\Gamma_{LO}$ denotes the linewidth of the zone-centered LO phonon. $k$ is the wavevector. $D$ represents the correlation length, which quantifies the distance that a phonon can propagate before being scattered by defects. Thus, $D$ is inversely correlated to the degree of SiC surface modification. By analyzing the spectra using this model, the continuum in the FLO3-4 and FLO2-4 patterns can be well reproduced by $D_{3-4} = 4.0 \pm 0.1$ nm and $D_{2-4} = 2.0 \pm 0.1$ nm (more details in Figs. 3c-d and Supplementary Note 4). Therefore, we assign the continuum in the Fano resonance to the confined LO phonon of modified SiC. The larger $D_{3-4}$ in the FLO3-4 pattern indicates a lower density of Si vacancies and less phonon scattering. The resulting spectral weight of the confined LO phonon is concentrated at the frequency of the discrete FLO phonon, facilitating the strong Fano resonance.

In addition to the close frequencies, another critical factor for the giant Fano resonance is the similar lifetimes of the FLO phonon (discrete) and the confined LO phonon (continuous), both of which are picoseconds. The importance of lifetime matching has been studied for Fano resonance in photonic systems[21,22]. For instance, in a metamaterial/protein structure, strong Fano resonance takes place when the lifetime of metamaterial scattering coincides with that of protein vibrational modes.[21] In solid-state materials, it has been demonstrated that the lifetime of coherent acoustic phonons can be tuned into a few nanoseconds to promote their coupling to electron-hole pairs[34]. Furthermore, theoretical models show that the confined phonon in nanostructures exhibits a long coherence time close to its lifetime[35]. In our study, the phonon coherence supported by the confined LO phonon of SiC can promote interference between continuous and discrete states, leading to strong Fano resonance.

Based on the frequency analysis of the continuum, we can rule out several other collective excitations as candidates for the continuum, such as the phonon-plasmon coupled mode[36] and plasmon polariton of SiC, both occurring at frequencies above 965 cm$^{-1}$. Surface plasmon polariton (SPP)[37] and surface phonon polariton (SPhP)[38] are unlikely to contribute to the continuum as the phonon intensities in our experiments display negligible dependence on the polarization of incident light (Fig. S10). Furthermore, our investigation reveals that the frequency of the FLO phonon does not depend on the incident power (Fig. S11). This rules out plasmons, whose frequency should change with the modulation of carrier density induced by incident power[39]. A comprehensive analysis of potential candidates for the continuum has been detailed in Supplementary Note 5.

**Fano interference modified by temperature**

The assignment of the scattering continuum to the confined LO phonon is further supported by the temperature dependence of phonon asymmetry. As depicted in Figures 4a-b, we observed a blue shift in the frequency of the FLO modes with decreasing temperature, akin to the symmetric phonon modes (Figs. S12-13). As shown in Fig. 4c, the asymmetry parameter $|1/q|$ of the FLO4 mode in the FLO2-4 pattern decreases from 1.0 to 0.5 as the temperature drops from room temperature to below 200 K. This temperature dependence is in agreement with the phonon-based Fano resonance reported for $KTa_{1-x}Nb_xO_3$, where the Fano resonance is also attributed to the interference between phonons[24]. In $KTa_{1-x}Nb_xO_3$, the measured $|1/q|$ decreases with lower temperatures similar to our results. In contrast, $|1/q|$ increases with reduced temperatures when the continuum is constituted by electronic transitions. For example, from room temperature to 4 K, the $|1/q|$ transitions from < 0.1 to 1.1 for the Fano resonance between Weyl fermions and phonons in TaAs[40]. In Fano resonance, $|1/q| = \left|\frac{V_E T_c}{T_d}\right|$ in which $T_c$ and $T_d$ denote the transition matrix elements of the continuum and discrete states[2], respectively, which are related to the respective phonon intensity. $|V_E|$ denotes the coupling between the continuum and discrete states. When the electronic transition serves as the continuum, reduced thermal excitation at lower temperatures leads to decreased Coulomb screening[40], contributing to a larger $T_c$. However, when phonons act as the continuum, $|V_E|$ plays a more important role on the temperature dependence, especially when $T_c$ and $T_d$ are on the same scale in the case of FLO2-4 pattern. The phonon-phonon interaction due to phonon anharmonicity at higher temperatures[41-43] can contribute to a larger $|V_E|$, thus explaining the temperature dependence of $|1/q|$ in Fano resonance involving phonons. For the FLO3-4 pattern, the FLO4 phonon with total antiresonance show a 50% decrease in $|1/q|$ when the temperature increases from 4 K to 295 K (Fig. 4d). This behavior suggests the dominance of $T_c$ over $|1/q|$ for the FLO3-4 pattern, which is further explained below.

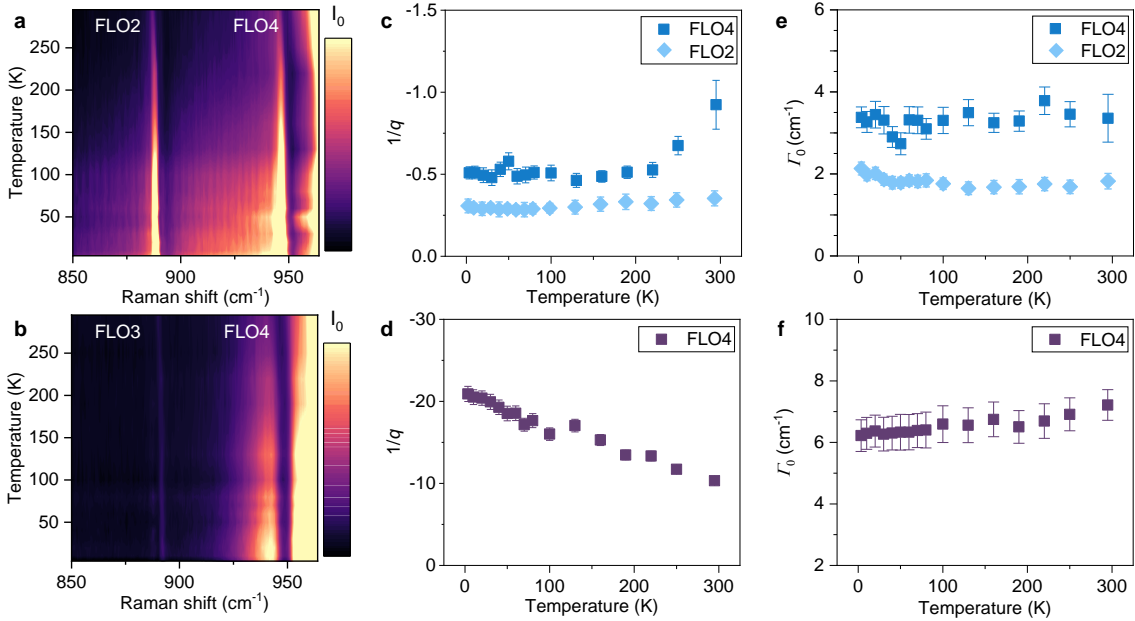

**Fig. 4**. Temperature dependence of Fano resonance. Raman intensity maps of **a**, FLO2-4 and **b**, FLO3-4 patterns at different temperatures. The fitted $1/q$ of the FLO2 and FLO4 modes as a function of temperature for **c**, FLO2-4 and **d**, FLO3-4 patterns. The fitted $\Gamma_0$ of the FLO2 and FLO4 modes as a function of temperature for **e**, FLO2-4 and **f**, FLO3-4 patterns. The error bars represent the 95% confidence interval obtained from fitting.

Besides the asymmetry, Fano resonance significantly modifies the FLO phonon lifetime. As shown in Fig. 4e, the linewidth of the FLO phonon does not have a prominent dependence on the temperature from 4 K to 295 K, while other symmetric phonons exhibit a 15% increase in linewidth (Fig. S14). At 4 K, the linewidth of the FLO4 mode in the FLO2-4 pattern is 3.4 cm$^{-1}$ (Fig. 4e), which is approximately half the linewidth of the FLO4 mode in the FLO3-4 pattern, that is, 6.2 cm$^{-1}$ (Fig. 4f). For optical phonon modes, the linewidth $\Gamma_0$ is related to the inverse of phonon lifetime $\tau$ ($\Gamma_0 \sim 1/\tau$)[44]. The different linewidths for the two Fano patterns suggest that the lifetime of the FLO4 mode in the FLO2-4 pattern ($\tau$ = 9.8 ps and $|1/q|$ = 0.5) is around twice that in the FLO3-4 pattern ($\tau$ = 5.4 ps and $|1/q|$ = 30.8). This difference indicates that the lifetime of the dressed phonon ($\tau$) is negatively correlated with the level of Fano resonance ($|1/q|$), as phonon-phonon interaction modulates the phonon self-energy and reduces the phonon lifetime[45-47]. Furthermore, we analyzed the transition probability of the continuous state for the two Fano patterns, leveraging the extracted values of $\tau$ and $|1/q|$. Ugo Fano's seminal paper[2] established the relationships $1/\tau \propto |V_E|^2$ and $|1/q|^2 = \left|\frac{V_E T_c}{T_d}\right|^2$. Hence, the ratio between the transition probability of the continuous states and the discrete state denoted as $\frac{|T_c|^2}{|T_d|^2}$, is proportional to the product of $\tau$ and $|1/q|^2$. For the FLO4 mode at 4 K, the $\tau \times |1/q|^2$ values are around 2.45 and 2.38×10$^3$ for the FLO2-4 and FLO3-4 patterns, respectively. This disparity implies that $\frac{|T_c|^2}{|T_d|^2}$ for the FLO3-4 pattern is roughly three orders of magnitude higher compared to the FLO2-4 pattern, justifying the conclusion that $T_c$ dominates $|1/q|$ for the FLO3-4 pattern.

**Tuning Fano resonance by 2D metals**

We have shown that the presence of 2D Ag is essential for achieving strong phonon-based Fano resonance in SiC. We will further demonstrate that the packing density of 2D Ag significantly influences the dependence of Fano resonance on different excitations (Figs. 5a-b), which offers an additional degree of tunability. The 2D Ag exhibits two distinct packing configurations as shown by the STEM images in Figs. 5c-d (corresponding atomic structure in Figs. 5e-f). In the case of the 4/3 packing, the unit cell contains four Ag atoms and three Si atoms as illustrated in Fig. 5e and Fig. S15. Conversely, the 3/3 packing comprises three Ag atoms and three Si atoms in the unit cell, with each Ag atom positioned atop one Si atom (Fig. 5f). Fourier transform analysis of the STEM images displays a lateral spacing of 0.150 nm between Ag atoms in the 4/3 packing (Fig. 5c). In contrast, the lateral spacing between Ag atoms in the 3/3 packing is 0.277 nm (Fig. 5d), which closely aligns with the 0.278 nm lateral spacing between Si atoms. These distinct packing configurations result in different dependencies on the excitation wavelength. For instance, as shown for 4/3 packing in Fig. 5a, the Fano resonance is most prominent under 532 nm excitation, while it remains absent under other excitations. In contrast, the 2D Ag sample with 3/3 packing does not exhibit Fano resonance under 532 nm excitation (Fig. 5b). Instead, the asymmetric FLO modes and scattering background in 2D Ag sample with 3/3 packing are nonnegligible when excited by 488 nm and 633 nm lasers (Fig. 5b).

To understand the relationship between the atomic structure of 2D Ag and the wavelength dependence of Fano resonance, we performed differential reflectance measurements on the 2D Ag sample characterized by 4/3 packing and 3/3 packing. The acquired differential reflectance spectra of 2D Ag unveil the distinct absorption features associated with packings. Specifically, the 4/3 packing displays optical absorption at 560 nm, close to the preferred excitation of 532 nm (Fig. 5a). Whereas the 3/3 packing exhibits optical absorption at 470 nm and minor absorption at 665 nm (Fig. 5g), which aligns with the Fano resonance observed under 488 nm and 633 nm excitations (Fig. 5b). These optical absorptions are attributed to the interband transitions in 2D Ag, which are inherently governed by the specific atomic configurations. The calculated imaginary part of the dielectric function in Fig. 5h (see Methods) successfully reproduces the optical absorption characteristics of 2D Ag with contrasting packings (Fig. 5g). The agreement between the optical absorption of 2D Ag and the preferred excitation suggests that the Fano resonance is supported by the resonant Raman scattering process.

The resonant Raman scattering is described by third-order perturbation theory, where the Raman intensity is written as

$$I(\omega_v) \propto \frac{1}{(E_L - E^{mi} - i\gamma)(E_L - E^{m'i} - \hbar\omega_v - i\gamma)} \quad (3)$$

in which $E_L$ is the laser energy, $E^{m(m')i}$ is the energy difference between state $i$ and state $m(m')$. More details are in Supplementary Note 6. Two resonant conditions can occur that enhance the Raman intensity (Fig. 5i insets), including incident resonant scattering, where incident photons resonate with interband transitions ($E_L = E_A$), and scattered resonant scattering, where incident photons differ from interband transitions by phonon energy $\hbar\omega$ ($E_L = E_A + \hbar\omega$). Such resonant conditions pictured here are consistent with the observations in the 2D Ag sample with 4/3 packing. For example, the optical absorption energy of the 2D Ag sample with 4/3 packing differs from the energy of the 532 nm laser by 116.5 meV (940 cm$^{-1}$), which is close to the frequency of the confined LO phonon (945 cm$^{-1}$). Thus, the strong intensity can be attributed to resonant Raman scattering assisted by a phonon. Since Fano asymmetry $|1/q|$ is related to phonon intensity through $\left|\frac{V_E T_C}{T_d}\right|$, the multiple scattering processes for the confined LO phonon may lead to a larger increase in $T_c$ and consequently a higher $|1/q|$. Importantly, the observed strong Fano resonance is unique to 2D metals, as it cannot be reproduced by thin-film Ag due to the lack of absorption in the visible regime (Fig. S16) and the inability to modify interfacial SiC to support the continuum states. This distinction highlights the fundamental differences between the optical properties of 2D metals and conventional metal thin films. The corroborated experimental and theoretical findings demonstrate that the atomic registry of 2D Ag governs its optical absorption and, in turn, gives rise to the excitation-dependent Fano resonance.

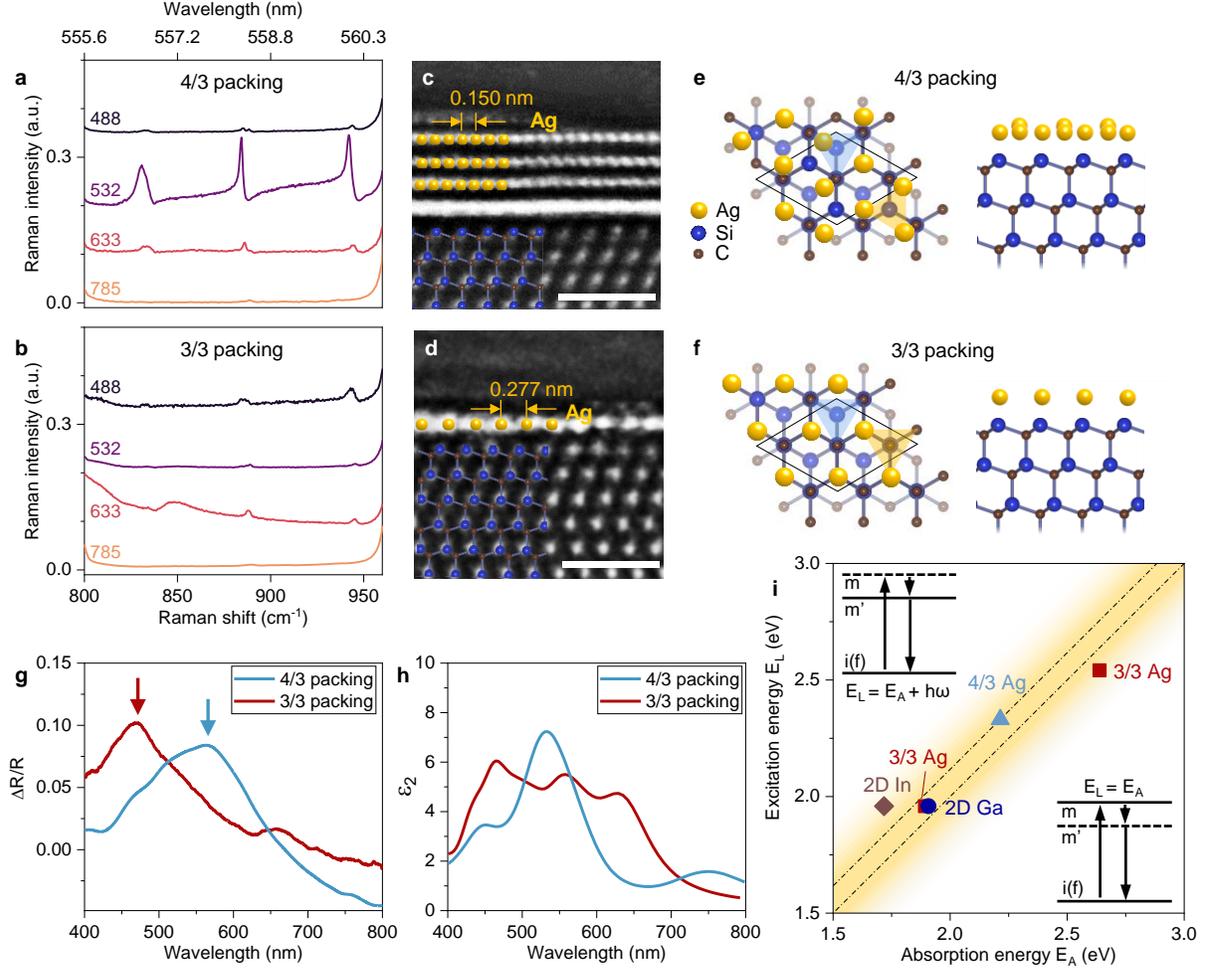

**Fig. 5**. The dependence of Fano resonance on the optical absorption of 2D metals. Raman spectra of 2D Ag with **a**, 4/3 packing and **b**, 3/3 packing under 488 nm, 532 nm, 633 nm, and 785 nm excitations. The spectra are normalized by the optical phonon at 788 cm$^{-1}$ and shifted along the y-axis. Cross-sectional STEM images of **c**, 4/3 packing and **d**, 3/3 packing. The scale bar is 1 nm. The atomic structures of 2D Ag/SiC **e**, with 4/3 packing and **f**, 3/3 packing. **g**, Differential reflectance $\frac{\Delta R}{R}$ of 2D Ag with different packings. $\frac{\Delta R}{R} = \frac{R(\omega) - R_0(\omega)}{R_0(\omega)}$, where R(ω) is the reflectance of graphene/2D Ag/SiC and R$_0$(ω) is the reflectance of graphene/SiC. Graphene exhibits minimal absorption features in this wavelength range (Fig. S19). **h**, The calculated ε$_2$ for 2D Ag/SiC with 4/3 and 3/3 packings. **i**, The correlation between the excitation energy and the optical absorption for different 2D metals. The excitation energy is when the Fano modes have the strongest intensity for each type of metal. Insets are the illustration of incident and scattered resonant Raman scattering.

Beyond 2D Ag, other 2D metals also enable Fano resonance for different photon energies as summarized in Fig. 5i. For example, Fano resonance has been observed in 2D gallium (2D Ga) for photon energy of 1.96 eV and 2D indium (2D In) for photon energies of 1.96 eV and 1.58 eV, which agrees with the optical absorption of 2D Ga at 1.91 eV and 2D In at 1.72 eV (Figs. S17-18). Figure 5i summarizes the correlation between the preferred excitation and optical absorption for different types of 2D metal. The data points are distributed along the resonance window, signifying that the resonant Raman scatterings induced by 2D metal are essential for the activation of Fano resonance.

**Ultrasensitive molecule detection based on Fano resonance**

As we have demonstrated, the activation of the Fano resonance strongly depends on resonant Raman scattering supported by the 2D metal. This resonance is sensitive to excitation wavelength, with the 532 nm laser being the most effective one among those we tested for 2D Ag with 4/3 packing. Therefore, any analyte that affects the light absorbed by the 2D metal may change the Fano features. For example, the introduction of Rhodamine 6G (R6G) molecules is likely to influence the 2D Ag absorption and Fano resonance, because the R6G fluorescence at 565 nm closely aligns with the optical absorption of 2D Ag at 560 nm (Fig. 6a). Thus, we hypothesize that changes in Fano features can serve as a sensitive indicator for detecting low concentrations of molecules. To demonstrate the possibility of molecule sensing, we deposited R6G molecules of different concentrations onto 2D Ag samples with the FLO2-4 pattern. As shown in Fig. 6b and Fig. S20, the intensity of the discrete FLO4 mode increases when higher concentrations of R6G molecules are adsorbed onto 2D Ag samples under 532 nm excitation. Further examination by fitting the intensity and $1/q$ of the FLO4 mode in Figs. 6c-d shows the dependence of Fano resonance on R6G concentrations, where the intensity and $|1/q|$ are monotonically correlated with the R6G concentration. A similar trend was also observed for the FLO3-4 pattern (Fig. S21). R6G with concentrations down to $10^{-12}$ M can be detected by monitoring the change of Fano intensity and $|1/q|$ compared to the pristine 2D Ag sample. This limit of detection of $10^{-12}$ M is at the single-molecule level (analysis of the single-molecule detection is in Supplementary Note 9), which is comparable to the most sensitive detection limit reported for surface-enhanced Raman scattering of R6G[48,49]. Remarkably, with a single R6G molecule, we can already observe significant changes in the Fano mode as seen in Figs. 6c-d: as large as 43% change of intensity and 87% change of $|1/q|$ compared to pristine 2D Ag sample. This shows the exquisite sensitivity of Fano modes to the surface molecules.

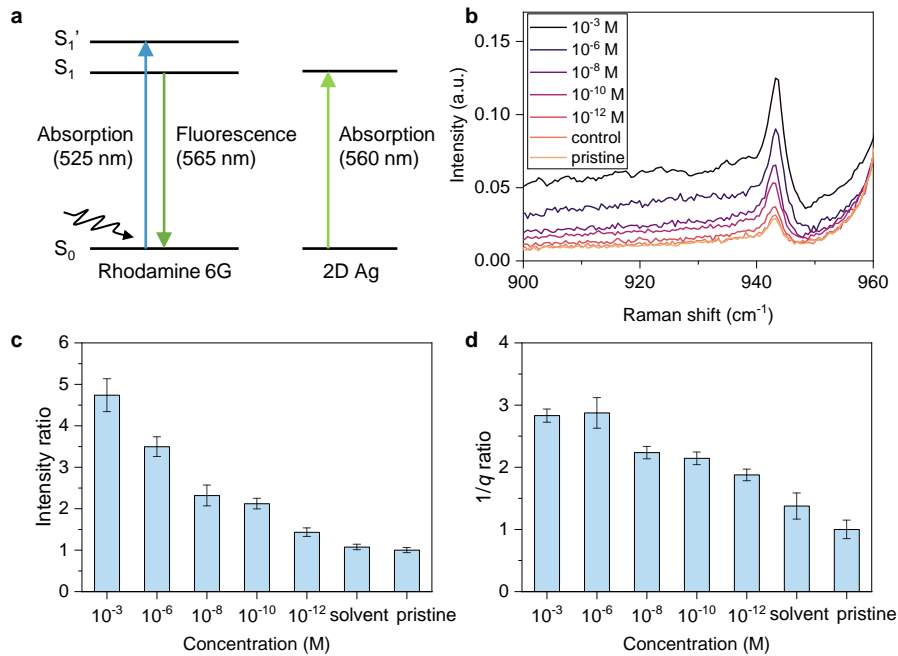

**Fig. 6**. Single-molecule detection based on Fano resonance. **a**, Electronic state diagram of R6G and 2D Ag with 4/3 packing. **b**, Raman spectra of 2D Ag samples with the FLO2-4 pattern treated with R6G of different concentrations under 532 nm excitation. "Solvent" refers to the 2D Ag sample treated with solvent only, and "pristine" is as-grown 2D Ag sample. The concentration of $10^{-12}$ M corresponds to having a single molecule under the laser beam. **c**, The intensity ratio and **d**, $1/q$ ratio of the FLO4 mode for 2D Ag sample treated with R6G of different concentrations to the pristine 2D Ag sample. The error bars represent the standard deviation of five different locations.

The observed change in Fano modes upon introducing R6G can be explained by the matched wavelengths between the R6G fluorescence and 2D metal absorption, which affects the Fano resonance. 2D Ag has an absorption peak at 560 nm (Fig. 5g). In the absence of R6G molecules, 2D Ag is directly excited by the monochromatic 532 nm laser, however, this light-matter interaction is less effective because the excitation wavelength mismatches with the absorption peak of 2D Ag. When R6G is present, the molecule efficiently absorbs the 532 nm photon from the laser (R6G absorption peak is around 525 nm) and subsequently emits broadband fluorescence photon within 540-600 nm with a peak at 565 nm (see Fig. S22). This fluorescence closely aligns with the broad optical absorption of 2D Ag (500-600 nm with a peak at 560 nm) better than the incident 532 nm laser (Fig. 6a), leading to a resonance between the R6G fluorescence and the optical absorption of 2D Ag. Therefore, the presence of R6G creates a sequential wavelength matching—between the laser and R6G absorption (532/525 nm) and between R6G fluorescence and metal absorption (565/560 nm), which facilitates the resonant energy transfer within the heterostructure. Eventually, this enhances the light effectively absorbed by the 2D metal and results in a pronounced change in the Fano modes.

The mechanism of such resonant energy transfer bears similarities to the Förster resonance energy transfer (FRET) process, where energy is transferred from a donor molecule to an acceptor molecule through dipole-dipole interaction. The efficiency of energy transfer depends on the spectral overlap between the donor emission and the acceptor absorption. It has been widely used for sensing and imaging of small chemical and biological molecules[50]. Similarly, in our system, the wavelength matching between the donor (R6G) and the acceptor (2D metal) facilitates the energy transfer that obviously modifies the Fano resonance.

**Conclusions**

In this work, we have demonstrated the remarkable phenomenon of giant Fano resonance in SiC enabled by the intercalation of 2D Ag. The discrete and continuous states contributing to Fano resonance originated from the FLO phonon and confined LO phonon of SiC at the 2D metal/SiC interface, respectively, which provides the frequency and lifetime matchings between scattering pathways. Notably, the FLO4 mode in the FLO3-4 pattern exhibited Fano resonance with $|1/q|$ two orders of magnitude higher than prior reported values. We explained the diverse Fano patterns by the specific surface termination of SiC formed after metal intercalation. The level of Fano resonance can be effectively tuned by creating defects in SiC through metal intercalation, with an effect similar to neutron scattering. We have also established that 2D Ag plays a vital role in inducing Fano resonance through resonant Raman scatterings. Such resonant Raman scattering greatly enhances the scattering intensity of phonon modes when incident photon energy matches the optical absorption of 2D metals or differs from that by phonon energies, which accounts for the extremely strong Fano resonance of the FLO4 mode. In addition to 2D Ag, we found that the intercalation of other 2D metals, such as Ga and In, also enables Fano resonance in SiC. This indicates the potential for manipulating Fano resonance by engineering the optical properties of intercalated 2D metals, including intercalating 2D alloys with tunable chemical compositions[51]. As another possible route to achieve resonance effect, near-field infrared photons can be used as the excitation, which may further improve the scattering efficiency of the phonon modes[52].

The significance of the phonon-related Fano resonance lies in several aspects. Firstly, the sensitivity of the Fano resonance pattern to the SiC terminations and defect levels can serve as a non-destructive and rapid probe for analyzing the atomic configurations of SiC. Furthermore, the resonant Raman scattering enabled by 2D metals sheds light on engineering quantum interference between bosonic quasiparticles at the nanoscale level. Leveraging this surface sensitivity, we further demonstrated that the asymmetric Fano modes can be used as a signature for ultrasensitive molecule detection at the single-molecule level. In the long run, the giant Fano resonance can offer opportunities for ultrasensitive optical detection, high-efficiency thermoelectrics, and superconductor devices, the operations of which are strongly influenced by phonons.

**Methods**

**Sample preparation**: Epitaxial Graphene (EG) is synthesized via sublimation of silicon from the (0001) plane of semi-insulating 6H-SiC (II-VI Inc.) at 1,800 °C, 93325.7 Pa, and 500 cm$^3$ STP/min Ar flow, for 30 min. EG is then plasma treated using a Tepla M4L plasma etch tool, using 150 cm$^3_{STP}$/min of O$_2$ and 50 cm$^3_{STP}$/min of He, under a pressure of 66.6 Pa and power of 50 W for 60 s. Metal intercalation was performed using a Thermo Scientific Lindberg/Blue M Mini-Mite tube furnace fitted with a 2.54 cm outer diameter quartz tube. A custom-made alumina crucible from Robocasting Enterprises was used to hold 1×1 cm$^2$ EG/SiC substrates. The EG side is placed downwards inside the crucible. Then, 75 mg of silver powder (Sigma Aldrich, 99.999 %) was placed in the crucible directly beneath the EG/SiC substrate. The crucible with EG/SiC and the silver powder was then loaded into the tube furnace and evacuated to be around 0.6 Pa. The tube was then pressurized to 66661.2 Pa with Ar. The furnace was then heated to 950 °C with a ramp rate of 50 °C min$^{-1}$ and an Ar flow of 50 cm$^3_{STP}$/min. The furnace was held at the growth temperature for 1 hour, then cooled to room temperature. For molecule detection, solutions of R6G in deionized water with concentrations ranging from $10^{-3}$ to $10^{-13}$ M were prepared. 50 microliters of the solution were drop cast onto a 5 mm × 5 mm 2D Ag. After 1 hour, the sample was rinsed with deionized water to remove excess molecules and dried with a nitrogen gun before Raman measurements.

**Optical measurement**: Raman measurements at room temperature were performed on a Horiba LabRAM Raman spectrometer equipped with the 1800/mm grating. The 100× objective with an NA of 0.9 was used to focus the laser on the sample. The ultra-low frequency module was added for measuring the low-frequency Raman modes under a 532 nm laser. For low-temperature Raman measurement, the sample was mounted inside a closed-cycled magneto-optical cryostat and immersed in He exchange gas. Low-temperature Raman spectra were measured using a triple grating spectrometer and in the backscattering geometry, where a 532 nm laser was used to excite an area of approximately 1 µm$^2$ on the sample. The laser power was kept at ≤500 µW to avoid local heating on the sample. The micro-reflection spectroscopy was measured on a home-built spectroscopy system using Laser-Driven Light Sources as the broadband white light source. We obtained the differential reflectance of 2D metals using EG on SiC as the reference. The graphene overlayer does not affect the analysis of absorption (Fig. S19). Details of the analysis are included in Supplementary Note 7.

**Scanning Transmission electron microscopy**: Thin cross-sectional STEM specimens were prepared by using the focused ion beam (FIB, FEI Helios 660) lift-out technique. A thick layer of protective amorphous carbon layer was deposited over the region of interest then Ga+ ions (30kV then stepped down to 1kV to avoid ion beam damage to the sample surface) were used in the FIB to make the samples electron transparent. Then the local microstructures of the cross-sectional samples were observed by FEI Titan3 G2 double aberration-corrected microscope at 300kV. All scanning transmission electron microscope (STEM) images were collected by using a high-angle annular dark field (HAADF) detector which had a collection angle of 52 mrad to 253 mrad. EDS elemental maps of the sample were collected by using a SuperX EDS system under STEM mode.

**First-principles calculations**: All DFT calculations were performed using the Vienna Simulation Package (VASP)[53]. We used the Perdew-Burke-Ernzerhof (PBE) parametrization of the generalized gradient approximation exchange-correlation functional[54], with additional corrections for van der Waals interactions using Grimme's semiempirical DFT-D3 correction scheme[55]. Other calculation parameters included the following: projector augmented wave pseudopotentials[56] for all elements, a plane-wave expansion energy cutoff of 500 eV, a relaxation force threshold of 10 meV/Å, and a 12×12×1 $k$-point grid for Ag/SiC in the 3/3 and 4/3 packing structures, as well as for bulk SiC. Structures prepared for phonon calculations were relaxed with a tighter force threshold of 5 meV/Å. Since bulk SiC is polar, correct LO-TO splitting in its phonon dispersion is achieved by applying a non-analytical correction (constructed from Born effective charge tensors) to the calculated dynamical matrix[57]. All interband dielectric functions were calculated at the independent particle level. A scissors shift of +0.5 eV was extracted from the comparison of PBE-level and HSE06-level[58] calculations for monolayer Ag/SiC and was applied to the optical spectra of the 3/3 and 4/3 packing structures so that all energy intervals involved in interband transition are increased by 0.5 eV to better compare with experiments. More details can be found in Supplementary Note 8.

## Author Contributions

K.Z., M.T.W., and A.J. performed Raman measurement at room temperature with input from S.V.R. K.Z. measured the differential reflectance spectra. R.A.M., A.J., C.D., and T.B. synthesized the EG and 2D metal samples. Y.W. performed first-principles calculations. T.T.M. conducted the low-temperature Raman measurement under the supervision of A.R.H. K.W. acquired the cross-sectional STEM images. S.H., J.R., and V.H.C. supervised the project. K.Z. performed the data analysis and wrote the manuscript with input from all authors.

## Acknowledgment

K.Z., R.N., S.V.R., V.H.C., J.R., and S.H. acknowledge the support from National Science Foundation (NSF) Award No. DMR-2011839 (through the Penn State MRSEC–Center for Nanoscale Science). K.Z. and S.H. also acknowledge the support from NSF Award No. ECCS-2246564 and ECCS-1943895. The co-authors acknowledge the use of the Penn State Materials Characterization Lab. K.Z. and S.H. acknowledge the helpful discussion on plasmons with Tony Low. K.Z. acknowledges Liming Gao for the useful discussion on the confinement model fitting. Y.W. acknowledges departmental startup funds from the University of North Texas. NIST Commercial Disclaimer: Certain commercial equipment, instruments, or materials are identified in this manuscript in order to specify the experimental procedure adequately. Such identification is not intended to imply recommendation or endorsement by the National Institute of Standards and Technology, nor is it intended to imply that the materials or equipment are necessarily the best available for the purpose.

## Data Availability

The data supporting the findings of the study are given in the figures in the main text and the Supplementary Information. Additional raw data are available from the corresponding authors upon reasonable request.

## References

1. C. L. Degen, F. Reinhard & P. Cappellaro. Quantum sensing. *Reviews of Modern Physics* **89**, 035002, (2017).
2. U. Fano. Effects of configuration interaction on intensities and phase shifts. *Physical Review* **124**, 1866-1878, (1961).
3. B. Luk'yanchuk, N. I. Zheludev, S. A. Maier, N. J. Halas, P. Nordlander, H. Giessen & C. T. Chong. The Fano resonance in plasmonic nanostructures and metamaterials. *Nature Materials* **9**, 707-715, (2010).
4. M. F. Limonov. Fano resonance for applications. *Adv. Opt. Photon.* **13**, 703-771, (2021).
5. D. Yoon, D. Jeong, H.-J. Lee, R. Saito, Y.-W. Son, H. C. Lee & H. Cheong. Fano resonance in Raman scattering of graphene. *Carbon* **61**, 373-378, (2013).
6. E. H. Hasdeo, A. R. T. Nugraha, M. S. Dresselhaus & R. Saito. Breit-Wigner-Fano line shapes in Raman spectra of graphene. *Physical Review B* **90**, 245140, (2014).
7. T. Mitani, S.-i. Nakashima, K. Kojima, T. Kato & H. Okumura. Determination of carrier concentration by Fano interference of Raman scattering in heavily doped n-type 4H-SiC. *Journal of Applied Physics* **112**, (2012).
8. S. Witte, R. T. Zinkstok, W. Ubachs, W. Hogervorst & K. S. E. Eikema. Deep-ultraviolet quantum interference metrology with ultrashort laser pulses. *Science* **307**, 400-403, (2005).
9. J. Liu, X. Huang, F. Wang & W. Hong. Quantum interference effects in charge transport through single-molecule junctions: detection, manipulation, and application. *Accounts of Chemical Research* **52**, 151-160, (2019).
10. L. B. Ioffe, V. B. Geshkenbein, M. V. Feigel'man, A. L. Fauchère & G. Blatter. Environmentally decoupled sds-wave Josephson junctions for quantum computing. *Nature* **398**, 679-681, (1999).
11. M. F. Limonov, M. V. Rybin, A. N. Poddubny & Y. S. Kivshar. Fano resonances in photonics. *Nature Photonics* **11**, 543-554, (2017).
12. Z. Chen, J.-W. Luo & L.-W. Wang. Revealing angular momentum transfer channels and timescales in the ultrafast demagnetization process of ferromagnetic semiconductors. *Proceedings of the National Academy of Sciences* **116**, 19258-19263, (2019).
13. B. Ziaja, N. Medvedev, V. Tkachenko, T. Maltezopoulos & W. Wurth. Time-resolved observation of band-gap shrinking and electron-lattice thermalization within X-ray excited gallium arsenide. *Scientific Reports* **5**, 18068, (2015).
14. S. D. M. Brown, A. Jorio, P. Corio, M. S. Dresselhaus, G. Dresselhaus, R. Saito & K. Kneipp. Origin of the Breit-Wigner-Fano lineshape of the tangential G-band feature of metallic carbon nanotubes. *Physical Review B* **63**, 155414, (2001).
15. E. H. Hasdeo, A. R. T. Nugraha, K. Sato, M. S. Dresselhaus & R. Saito. Electronic Raman scattering and the Fano resonance in metallic carbon nanotubes. *Physical Review B* **88**, 115107, (2013).
16. J. Coulter, G. B. Osterhoudt, C. A. C. Garcia, Y. Wang, V. M. Plisson, B. Shen, N. Ni, K. S. Burch & P. Narang. Uncovering electron-phonon scattering and phonon dynamics in type-I Weyl semimetals. *Physical Review B* **100**, 220301, (2019).
17. K. Zhang, T. Wang, X. Pang, F. Han, S.-L. Shang, N. T. Hung, Z.-K. Liu, M.

Li, R. Saito & S. Huang. Anisotropic Fano resonance in the Weyl semimetal candidate LaAlSi. *Physical Review B* **102**, 235162, (2020).
18. Q.-H. Tan, Y.-J. Sun, X.-L. Liu, Y. Zhao, Q. Xiong, P.-H. Tan & J. Zhang. Observation of forbidden phonons, Fano resonance and dark excitons by resonance Raman scattering in few-layer $WS_2$. *2D Materials* **4**, 031007, (2017).
19. S. Das, S. Prasad, B. Chakraborty, B. Jariwala, S. Shradha, D. V. S. Muthu, A. Bhattacharya, U. V. Waghmare & A. K. Sood. Doping controlled Fano resonance in bilayer 1T′-$ReS_2$: Raman experiments and first-principles theoretical analysis. *Nanoscale* **13**, 1248-1256, (2021).
20. S. K. Saxena, P. Yogi, S. Mishra, H. M. Rai, V. Mishra, M. K. Warshi, S. Roy, P. Mondal, P. R. Sagdeo & R. Kumar. Amplification or cancellation of Fano resonance and quantum confinement induced asymmetries in Raman line-shapes. *Physical Chemistry Chemical Physics* **19**, 31788-31795, (2017).
21. C. Wu, A. B. Khanikaev, R. Adato, N. Arju, A. A. Yanik, H. Altug & G. Shvets. Fano-resonant asymmetric metamaterials for ultrasensitive spectroscopy and identification of molecular monolayers. *Nature Materials* **11**, 69-75, (2012).
22. A. B. Khanikaev, C. Wu & G. Shvets. Fano-resonant metamaterials and their applications. *Nanophotonics* **2**, 247-264, (2013).
23. V. Dwij, B. K. De, M. K. Gupta, R. Mittal, N. P. Lalla & V. Sathe. Reversible optical control of Fano resonance and domain configuration at room temperature in $BaTiO_3$. *Journal of Applied Physics* **131**, 053102, (2022).
24. Y. Wu, F. Liang, X. Wang, J. Wang, H. Yu & H. Zhang. Temperature dependent Raman spectroscopic study of Fano resonance in perovskite ferroelectric $KTa_{1-x}Nb_xO_3$ single crystal. *Opt. Mater. Express* **12**, 247-255, (2022).
25. H. J. Butler, L. Ashton, B. Bird, G. Cinque, K. Curtis, J. Dorney, K. Esmonde-White, N. J. Fullwood, B. Gardner, P. L. Martin-Hirsch, M. J. Walsh, M. R. McAinsh, N. Stone & F. L. Martin. Using Raman spectroscopy to characterize biological materials. *Nature Protocols* **11**, 664-687, (2016).
26. N. Briggs, B. Bersch, Y. Wang, J. Jiang, R. J. Koch, N. Nayir, K. Wang, M. Kolmer, W. Ko, A. De La Fuente Duran, S. Subramanian, C. Dong, J. Shallenberger, M. Fu, Q. Zou, Y.-W. Chuang, Z. Gai, A.-P. Li, A. Bostwick, C. Jozwiak, C.-Z. Chang, E. Rotenberg, J. Zhu, A. C. T. van Duin, V. Crespi & J. A. Robinson. Atomically thin half-van der Waals metals enabled by confinement heteroepitaxy. *Nature Materials* **19**, 637-643, (2020).
27. H. El-Sherif, N. Briggs, B. Bersch, M. Pan, M. Hamidinejad, S. Rajabpour, T. Filleter, K. W. Kim, J. Robinson & N. D. Bassim. Scalable characterization of 2D gallium-intercalated epitaxial graphene. *ACS Applied Materials & Interfaces* **13**, 55428-55439, (2021).
28. D. Momeni Pakdehi, P. Schädlich, T. T. N. Nguyen, A. A. Zakharov, S. Wundrack, E. Najafidehaghani, F. Speck, K. Pierz, T. Seyller, C. Tegenkamp & H. W. Schumacher. Silicon carbide stacking-order-induced doping variation in epitaxial graphene. *Advanced Functional Materials* **30**, 2004695, (2020).
29. H. El-Sherif, N. Briggs, B. Bersch, J. Robinson & N. Bassim. Observation

of 2D Si-vacancies filled by gallium intercalation of epitaxial graphene. *Microscopy and Microanalysis* **28**, 2528-2530, (2022).
30. T. Koyanagi, M. J. Lance & Y. Katoh. Quantification of irradiation defects in beta-silicon carbide using Raman spectroscopy. *Scripta Materialia* **125**, 58-62, (2016).
31. S. Nakashima, T. Mitani, J. Senzaki, H. Okumura & T. Yamamoto. Deep ultraviolet Raman scattering characterization of ion-implanted SiC crystals. *Journal of Applied Physics* **97**, 123507, (2005).
32. H. Richter, Z. P. Wang & L. Ley. The one phonon Raman spectrum in microcrystalline silicon. *Solid State Communications* **39**, 625-629, (1981).
33. I. H. Campbell & P. M. Fauchet. The effects of microcrystal size and shape on the one phonon Raman spectra of crystalline semiconductors. *Solid State Communications* **58**, 739-741, (1986).
34. T. Vasileiadis, H. Zhang, H. Wang, M. Bonn, G. Fytas & B. Graczykowski. Frequency-domain study of nonthermal gigahertz phonons reveals Fano coupling to charge carriers. *Science Advances* **6**, eabd4540.
35. Z. Zhang, Y. Guo, M. Bescond, M. Nomura, S. Volz & J. Chen. Assessing phonon coherence using spectroscopy. *Physical Review B* **107**, 155426, (2023).
36. H. Harima, S. i. Nakashima & T. Uemura. Raman scattering from anisotropic LO-phonon–plasmon– coupled mode in n-type 4H– and 6H– SiC. *Journal of Applied Physics* **78**, 1996-2005, (1995).
37. J. Lin, J. P. B. Mueller, Q. Wang, G. Yuan, N. Antoniou, X.-C. Yuan & F. Capasso. Polarization-controlled tunable directional coupling of surface plasmon polaritons. *Science* **340**, 331-334, (2013).
38. G. Zheng, L. Xu, X. Zou & Y. Liu. Excitation of surface phonon polariton modes in gold gratings with silicon carbide substrate and their potential sensing applications. *Applied Surface Science* **396**, 711-716, (2017).
39. H. Yu, Y. Peng, Y. Yang & Z.-Y. Li. Plasmon-enhanced light–matter interactions and applications. *npj Computational Materials* **5**, 45, (2019).
40. B. Xu, Y. M. Dai, L. X. Zhao, K. Wang, R. Yang, W. Zhang, J. Y. Liu, H. Xiao, G. F. Chen, S. A. Trugman, J. X. Zhu, A. J. Taylor, D. A. Yarotski, R. P. Prasankumar & X. G. Qiu. Temperature-tunable Fano resonance induced by strong coupling between Weyl fermions and phonons in TaAs. *Nature Communications* **8**, 14933, (2017).
41. E. Liarokapis, E. Anastassakis & G. A. Kourouklis. Raman study of phonon anharmonicity in $LaF_3$. *Physical Review B* **32**, 8346-8355, (1985).
42. X. Gu, Z. Fan, H. Bao & C. Y. Zhao. Revisiting phonon-phonon scattering in single-layer graphene. *Physical Review B* **100**, 064306, (2019).
43. D. Tristant, A. Cupo, X. Ling & V. Meunier. Phonon anharmonicity in few-layer black phosphorus. *ACS Nano* **13**, 10456-10468, (2019).
44. A. Togo, L. Chaput & I. Tanaka. Distributions of phonon lifetimes in Brillouin zones. *Physical Review B* **91**, 094306, (2015).
45. Y. Xu, J.-S. Wang, W. Duan, B.-L. Gu & B. Li. Nonequilibrium Green's function method for phonon-phonon interactions and ballistic-diffusive thermal transport. *Physical Review B* **78**, 224303, (2008).
46. L. Chaput, A. Togo, I. Tanaka & G. Hug. Phonon-phonon interactions in

transition metals. *Physical Review B* **84**, 094302, (2011).
47. A. Glensk, B. Grabowski, T. Hickel, J. Neugebauer, J. Neuhaus, K. Hradil, W. Petry & M. Leitner. Phonon lifetimes throughout the Brillouin zone at elevated temperatures from experiment and Ab Initio. *Physical Review Letters* **123**, 235501, (2019).
48. L. Qu, N. Wang, H. Xu, W. Wang, Y. Liu, L. Kuo, T. P. Yadav, J. Wu, J. Joyner, Y. Song, H. Li, J. Lou, R. Vajtai & P. M. Ajayan. Gold nanoparticles and g-$C_3N_4$-intercalated graphene oxide membrane for recyclable surface enhanced Raman scattering. *Advanced Functional Materials* **27**, 1701714, (2017).
49. L. Tao, K. Chen, Z. Chen, C. Cong, C. Qiu, J. Chen, X. Wang, H. Chen, T. Yu, W. Xie, S. Deng & J.-B. Xu. 1T′ transition metal telluride atomic layers for plasmon-free SERS at femtomolar levels. *Journal of the American Chemical Society* **140**, 8696-8704, (2018).
50. L. Wu, C. Huang, B. P. Emery, A. C. Sedgwick, S. D. Bull, X.-P. He, H. Tian, J. Yoon, J. L. Sessler & T. D. James. Förster resonance energy transfer (FRET)-based small-molecule sensors and imaging agents. *Chemical Society Reviews* **49**, 5110-5139, (2020).
51. S. Rajabpour, A. Vera, W. He, B. N. Katz, R. J. Koch, M. Lassaunière, X. Chen, C. Li, K. Nisi, H. El-Sherif, M. T. Wetherington, C. Dong, A. Bostwick, C. Jozwiak, A. C. T. van Duin, N. Bassim, J. Zhu, G.-C. Wang, U. Wurstbauer, E. Rotenberg, V. Crespi, S. Y. Quek & J. A. Robinson. Tunable 2D group-III metal alloys. *Advanced Materials* **33**, 2104265, (2021).
52. R. Hillenbrand, T. Taubner & F. Keilmann. Phonon-enhanced light–matter interaction at the nanometre scale. *Nature* **418**, 159-162, (2002).
53. G. Kresse & J. Furthmüller. Efficient iterative schemes for ab initio total-energy calculations using a plane-wave basis set. *Physical Review B* **54**, 11169-11186, (1996).
54. J. P. Perdew, K. Burke & M. Ernzerhof. Generalized gradient approximation made simple. *Physical Review Letters* **77**, 3865-3868, (1996).
55. S. Grimme, J. Antony, S. Ehrlich & H. Krieg. A consistent and accurate ab initio parametrization of density functional dispersion correction (DFT-D) for the 94 elements H-Pu. *The Journal of Chemical Physics* **132**, 154104, (2010).
56. P. E. Blöchl. Projector augmented-wave method. *Physical Review B* **50**, 17953-17979, (1994).
57. X. Gonze & C. Lee. Dynamical matrices, Born effective charges, dielectric permittivity tensors, and interatomic force constants from density-functional perturbation theory. *Physical Review B* **55**, 10355-10368, (1997).
58. A. V. Krukau, O. A. Vydrov, A. F. Izmaylov & G. E. Scuseria. Influence of the exchange screening parameter on the performance of screened hybrid functionals. *The Journal of Chemical Physics* **125**, 224106, (2006).